\documentclass[sigconf]{acmart}
\usepackage[utf8]{inputenc}
\usepackage{tikz}
\usepackage{amsmath}
\usepackage{filecontents}
\usepackage[flushleft]{threeparttable}
\usepackage{amsmath,amsfonts}

\usepackage{graphicx}
\usepackage{textcomp}

 \usepackage{url}

\usepackage{lipsum,tabularx}
\usepackage{pythonhighlight}
\usepackage{multicol}
\usepackage{multirow}

\usepackage{colortbl}
\usepackage{booktabs}
\usepackage{setspace}

\hypersetup{
  linkcolor={blue!70!black},
  citecolor={red!70!black},
  urlcolor={blue!70!black}
}
\def\Snospace~{\S{}}

\usepackage[T1]{fontenc}

\usepackage{algorithm}
\usepackage{algorithmicx}
\usepackage[noend]{algpseudocode}
\usepackage{balance}

\usepackage{bm}
\usepackage{fp}
\usepackage{siunitx}
\usepackage{amsthm}

\usepackage[labelfont=bf,font=small,skip=5pt]{caption}
\usepackage{subcaption}

\usepackage{comment}

\usepackage{fancyhdr}
\usepackage{tikz}

\usepackage{xspace}

\newcommand{\boxbeg}{
  \vspace{2px}
  \noindent\begin{tabular}{|l|}\hline
    \begin{minipage}{3.2in}
      \vspace{2px}
      \noindent
      }

      \newcommand{\boxend}{
      \vspace{2px}
    \end{minipage} \\ \hline
  \end{tabular}
  \vspace{-10pt}
}

\usepackage{algpseudocode}
\usepackage[htt]{hyphenat}

\usepackage{wrapfig}

\usepackage[scaled=.7]{beramono}
\usepackage{pifont}
\usepackage{mdframed}

\newcommand{\eg}[0]{e.g.}

\newcounter{finding}
\newcommand{\finding}[1]{\refstepcounter{finding}
    \vspace{0.5mm}
    \begin{mdframed}[linecolor=gray,roundcorner=12pt,backgroundcolor=gray!15,linewidth=3pt,innerleftmargin=2pt, leftmargin=0cm,rightmargin=0cm,topline=false,bottomline=false,rightline = false]
    \textbf{Finding \arabic{finding}:} #1
    \end{mdframed}
    \vspace{0.5mm}
}

\AtBeginDocument{%
  }

\setcopyright{acmcopyright}
\copyrightyear{2023}
\acmYear{2023}

\begin{document}

\newcommand{\ds}{\mbox{LibReady}\xspace}

\title{Rethinking Technology Stack Selection with AI Coding Proficiency }


\author{Xiaoyu Zhang}
\affiliation{%
  \institution{Xi'an Jiaotong University}
  \city{Xi'an}
  \country{China}
}
\email{joshingrain@gmail.com}

\author{Weipeng Jiang}
\affiliation{%
  \institution{Xi'an Jiaotong University}
  \country{China}}

\author{Shiqing Ma}
\affiliation{%
  \institution{University of Massachusetts, Amherst}
  \country{United States}}

\author{Qingshuang Bao}
\affiliation{%
  \institution{Xi'an Jiaotong University}
  \country{China}}

\author{Chenhao Lin}
\affiliation{%
  \institution{Xi'an Jiaotong University}
  \country{China}}

\author{Chao Shen}
\authornote{Chao Shen is the corresponding author.}
\affiliation{%
  \institution{Xi'an Jiaotong University}
  \country{China}}

\author{Tianlin Li}
\affiliation{%
  \institution{Nanyang Technological University}
  \country{Singapore}}

\author{Juan Zhai}
\affiliation{%
  \institution{University of Massachusetts, Amherst}
  \country{United States}}

\begin{abstract}
Large language models (LLMs) are now an integral part of software development workflows and are reshaping the whole process.
However, existing technology selection methods mainly focus on the inherent attributes of technologies, overlooking whether the LLM can effectively leverage the chosen technology.
Therefore, teams using LLM assistants risk choosing technologies that cannot be used effectively by LLMs, yielding high debugging effort and mounting technical debt.
We foresee a practical question in the LLM era, \textit{is a technology ready for AI-assisted development}?
In this paper, we first propose the concept, \textit{AI coding proficiency}, the degree to which LLMs can utilize a given technology to generate high-quality code snippets.
We then conduct the first large-scale empirical study examining AI coding proficiency across 170 third-party libraries and six LLMs.
Our findings reveal that libraries with similar functionalities can exhibit up to 84\% differences in the quality score of LLM-generated code. 
These gaps can be translated into real engineering costs and steer developer choices toward a narrow set of technologies, threatening technological diversity in the ecosystem.
We call on the community to integrate AI coding proficiency into technology selection frameworks and develop mitigation strategies, preserving competitive balance in AI-driven development.
\end{abstract}


\maketitle

\section{Introduction}\label{sec:intro}

Large language models (LLMs) have demonstrated impressive code generation capabilities, achieving performance that closes or even surpasses that of human developers~\cite{hou2025comparing,openai_gpt5_benchmarks}.
These models have been integrated into the daily workflows of developers worldwide, with recent reports indicating that over half (i.e., 51\%) of professional developers use AI tools in their daily practice~\cite{stackoverflow2025aicodingsurvey}.
This gives rise to new development styles such as `Vibe Coding' that heavily rely on LLM-based development~\cite{news_vibecoding,li2025prompting,gadde2025democratizing}, reshaping both the software ecosystem and development processes.

In the software development lifecycle, developers need to select appropriate programming languages, frameworks, and third-party libraries based on target requirements before coding, which is known as \textit{technology stack selection}~\cite{stratoflow2023sdlc}. 
For example, when building a web application, developers usually choose mainstream backend frameworks such as \texttt{Spring Boot}, \texttt{Django}, or \texttt{Express.js}, rather than building from scratch. 
Technology stack selection has a decisive impact on development efficiency, maintainability, and software quality~\cite{birrell1988practical,stratoflow2023sdlc}, as a good choice can reduce development costs by up to 78\%~\cite{TechDebtCost2025}, while an inappropriate stack can lead to project failure and economic losses~\cite{WrongTechnologyStack2025}.
Within this broader context, \textit{technology selection}, the choice of individual technologies or components within a stack, constitutes a key component of stack selection. 
Recognizing its importance, existing software engineering (SE) research has proposed evaluation frameworks and guidelines to support technology selection~\cite{kazman2000atam,bass2021software,kazman2001quantifying,iso25024,golden1989analytic}, systematically evaluating quality attributes such as maintainability, performance, and reliability to rank candidate technologies and facilitate optimal technology adoption~\cite{iso25024}.

However, the advent of LLMs necessitates a fundamental rethinking of technology selection, as solutions friendly to human developers may not be equally effective for workflows involving LLMs.
For instance, human users prefer readable natural language instructions, while scripted queries are easier for LLMs to parse accurately~\cite{beurer2023prompting}.
In software development, this mismatch between humans and models is becoming increasingly prominent.
Existing reports have shown that when generating code snippets using libraries popular for human developers, LLM could generate low-quality code snippets that call deprecated APIs or methods and may even contain syntax errors~\cite{codegen_for_llms2025,wang2025llms}.
Therefore, there is an urgent need to investigate this emerging challenge and rethink the evaluation framework for technology selection in the LLM era.
While existing research has focused on evaluating LLMs' internal code generation capability~\cite{jiang2024survey,peng2025coffe,tang2024gendercare}, to our knowledge, no work has systematically investigated the technology-dependent quality variations in LLM-generated code snippets, which is significant for technology selection in AI-assisted development.

To fill this gap, we firstly introduce a new perspective for technology selection in AI-assisted development in this paper, namely \textit{AI coding proficiency} (hereafter `AI proficiency'), a property of a technology (e.g., a third-party library) that quantifies the degree to which a given LLM can accurately understand and correctly invoke its APIs, thereby producing high-quality code snippets.
As shown in~\autoref{fig:moti}, even with similar functionality and popularity, technologies can exhibit substantial differences in AI proficiency.
The code snippets generated by GPT-4o using \texttt{Shapely} typically have lower quality than those using \texttt{GDAL} (up to 20.21\% on repeated runs of the same prompt), with some even exhibiting basic syntax errors.
The implications of AI proficiency are twofold.
At the engineering level, selecting a technology with low AI proficiency increases debugging effort and offsets LLM-driven productivity gains.
At the ecosystem level, proficiency gaps can trigger a model-driven `Matthew effect'~\cite{merton1968matthew}, where high-proficiency technologies attract disproportionate adoption and resources, suppressing technological diversity and raising potential security and legal risks~\cite{FTCAct,crandall2023antitrust}.
To systematically study AI coding proficiency, we construct an automated pipeline to build a dataset containing 170 third-party libraries across 61 scenarios.
Then, we adopt a multi-dimensional code quality assessment covering functional suitability, performance, maintainability, readability, and reliability, drawing on established code quality literature and the characteristics of LLM-generated code (i.e., code fragmentation and low context dependency).
Based on the dataset and assessment framework, we conduct a study on six representative LLMs.
Note that this study focuses on the Python ecosystem, where LLMs have demonstrated state-of-the-art code generation performance~\cite{twist2025llms,qing2025effibench}, and targets third-party libraries as the selection and management of these libraries is an important part of the Python ecosystem~\cite{wang2020watchman}.
This study aims to answer the following research questions (RQs).

\noindent
\(\bullet\)
\textit{\textbf{RQ1 (Proficiency Gap)}: How is the AI proficiency of different libraries on different LLMs? }
We evaluate the AI proficiency scores of 170 libraries on six LLMs and compare \ding{182} the AI proficiency differences between competing libraries with similar functionality and \ding{183} the proficiency differences of the same library on different LLMs.
The findings show that the difference in AI proficiency between competing libraries and models is common, and over 10\% of competing library pairs exhibit significant differences across different models.
In addition, the `winner' of competing libraries usually varies across LLMs.
We call on developers to incorporate AI proficiency into the technology selection framework and evaluate candidate libraries and LLMs simultaneously to obtain an effective combination to improve development efficiency and quality.


\noindent
\(\bullet\)
\textit{\textbf{RQ2 (Failure Pattern)}: What typical failure patterns do LLMs exhibit when generating low-quality code snippets? }
To gain a deeper understanding of the manifestations of low AI coding proficiency, we conduct an analysis of the code snippets marked with `low quality' (using the interquartile range method) in RQ1.
We identify eight typical failure patterns, where `Incorrect Functionality' and `Missing Edge Handling' are the most common failure patterns, accounting for over 40\% of all low-quality cases, directly endangering the safety and reliability of software development.

\noindent
\(\bullet\)
\textit{\textbf{RQ3 (Enhancement Strategy Exploration)}: Can existing prompting techniques improve the AI proficiency of libraries? }
From the perspective of model users, we evaluate three prompt engineering techniques.
Experiment results show that the prompting methods can improve the overall code quality and AI coding proficiency scores and reduce the AI proficiency gaps between libraries, which provides a feasible solution for mitigating potential risks caused by the proficiency differences.



The contributions of this paper are shown as follows.
\begin{itemize}
    \item We first propose and define AI coding proficiency of technologies, providing a novel perspective for technology selection in the LLM era.
    \item We construct a comprehensive evaluation methodology, including a multi-dimensional code quality assessment covering five dimensions and a pipeline to generate a dataset of 170 libraries across 61 scenarios, providing a solid foundation for future AI coding and technology selection research.
    \item We conduct the first large-scale empirical study to reveal and quantify the AI coding proficiency gaps of third-party libraries across six LLMs. Based on the six key findings, we call on developers to incorporate AI coding proficiency of different technologies on different models into technology selection to achieve high-quality code generation and efficient development.
    \item Our pipeline implementation, dataset, and the necessary results are available at~\cite{our_repo}.
\end{itemize}

\section{Background \& Related Work}
\label{sec:bg}

\subsection{LLM Code Generation}

LLMs, trained on large-scale code corpora, present new opportunities for SE.
They have demonstrated strong potential in a number of challenging tasks, such as automatic patch generation~\cite{ribeiro2023large}, code summarization~\cite{chai2022pyramid}, and repository-level code generation~\cite{zhang2023repocoder}.
At present, the industry and academia have released a large number of LLMs, which are accelerating the production of developers around the world~\cite{li2022competition,li2023starcoder}.
To measure the code generation performance of these LLMs, researchers have proposed a series of benchmarks~\cite{chen2021evaluating, austin2021program,lai2023ds,zhuo2025bigcodebench,fachada2025gpt} and use metrics like \texttt{pass@k} to evaluate whether the generated functions can pass all unit tests.
Recent work has further constructed benchmarks to evaluate the performance efficiency of the generated code snippets~\cite{huang2024effibench}.
However, real-world software development requires additional quality considerations~\cite{dou2026wrong}.
For example, LLM-generated code should be maintainable and reliable for collaborative development environments.
In this paper, we extend the perspective to a more comprehensive evaluation framework that assesses code quality across five dimensions, thereby describing how effectively an LLM can use a given technology in code generation.
By systematically measuring these quality dimensions when LLMs utilize specific third-party libraries in code generation, we quantify the AI coding proficiency of various libraries, aiming to provide insights for technology selection in LLM-assisted development.

\subsection{Technology Selection}
\label{s:tech_stack}

Technology stack selection involves choosing a suitable combination of languages, frameworks, and third-party libraries to meet software requirements, which is a foundational phase of the software development lifecycle~\cite{bourque2002guide,bass2021software}.
Technology selection, as a crucial component within this process, focuses on identifying the most suitable technology from functionally similar alternatives for a given task scenario~\cite{larios2020selecting,li2022exploring}.
The technology selection has long been treated as a multi-criteria decision process in which teams balance functional suitability, quality attributes, and cost under real-world constraints.
Researchers have proposed diverse methods like architecture evaluation methods (e.g., ATAM)~\cite{kazman2000atam,bass2021software,kazman2001quantifying}, multi-criteria decision-making methods~\cite{golden1989analytic,tzeng2011multiple}, and standards-based quality models~\cite{iso25024} that systematically balance quality attributes such as maintainability and reliability.

However, as software development shifts toward collaboration between LLM-driven AI tools and developers (as mentioned in~\autoref{sec:intro}), traditional approaches that only evaluate intrinsic attributes of technologies, overlooking whether the LLM can proficiently use the target technology to produce high-quality outputs.
To address this blind spot, we introduce \textit{AI coding proficiency} as a complementary perspective for technology selection, evaluating whether a technology can be effectively used by LLMs to produce high-quality code snippets~\cite{xia2019practitioners,rojpaisarnkit2024towards}.
Our study evaluates AI coding proficiency across different technologies and LLMs, revealing its significance as a new dimension for technology selection in the LLM era.

\subsection{Code Quality Evaluation}
\label{s:quality_eval}
Code quality is a multi-faceted concept that needs to be evaluated from multiple dimensions.
The prior ISO standard provides an authoritative and widely accepted taxonomy for code quality evaluation with several key dimensions like functional suitability~\cite{iso25024}.
Based on this standard, we review the evaluation methods of commonly used dimensions as follows.

{\bf Functional Suitability} represents the degree to which a system provides functions that fulfill the intended purpose and needs when used under specified conditions.
The related research on this dimension mainly evaluates the correctness of the generated code snippets.
For functional correctness, a common metric is \texttt{pass@k}, which quantifies the proportion of generated samples that successfully pass a predefined unit test suite~\cite{chen2021evaluating, austin2021program}.
However, this approach requires the manual construction of multiple unit tests (often dozens of unit tests) for each task, and the evaluation results heavily depend on the quality and coverage of these tests, thereby limiting its scalability in large-scale automated evaluation scenarios. 
Recently, with the advancement of LLMs' reasoning and comprehension capabilities, researchers have begun to widely use the LLM as a judge to evaluate and score code snippets statically~\cite{zhang2025codecriticbench,crupi2025effectiveness,xu2025unleashing,wang2025can}.
By prompting LLMs with detailed rules or structured checklists, these methods enable the systematic assessment of multiple attributes of code snippets, including functional correctness and complexity.

{\bf Performance Efficiency} relates to the resources the code snippets consume during execution.
In traditional SE, it is typically evaluated through a range of profiling techniques, including load testing of complete software systems and algorithmic complexity analysis~\cite{barna2011autonomic,draheim2006realistic,cormen2022introduction,mcgeoch2012guide}.
However, these system-level or theoretical methods are often unsuitable for the automated evaluation of the short and isolated snippets generated by LLMs.
Researchers have evaluated the efficiency of code snippets by profiling the average execution time and peak memory usage~\cite{huang2024effibench,chen2024survey}.
Recent work further explores the use of LLMs to analyze and estimate the time and space complexity of code snippets, thereby providing efficiency scores~\cite{zhang2025codecriticbench,chambon2025bigo}.

{\bf Maintainability} is the degree of effectiveness and efficiency with which a product can be modified.
This attribute is crucial for controlling software debugging and maintenance costs.
Traditional assessment combines code metrics with historical analysis (e.g., code churn) and inter-module dependency metrics~\cite{hassan2009predicting,nagappan2005use,briand2002unified}.
For LLM code snippets that lack historical and inter-module context, the evaluation focuses on intrinsic code properties, including the Halstead volume, cyclomatic complexity, and maintainability index~\cite{oman1992metrics,mccabe1976complexity,jin2023software}.
These classic metrics analyze the linearly independent paths and vocabulary in the source code, thereby calculating and evaluating the complexity and maintainability of the code snippets.
Recently, researchers have also explored using LLMs to analyze and evaluate the maintainability of code snippets based on predefined checklists~\cite{zhang2025codecriticbench}.

{\bf Usability} measures the ease with which users can understand and apply the code to achieve their goals.
Traditional evaluation is mainly human-centric, relying on methods like developer surveys, cognitive complexity studies, and think-aloud protocols to assess a library's or system's learning curve and ease of use~\cite{grossman2009survey,piccioni2013empirical}.
However, these methods are not scalable for automatically evaluating numerous code snippets.
Prior studies establish that readability is a key factor in a developer's ability to understand and use code snippets~\cite{letouzey2012managing,rauf2019systematic}.
Accordingly, existing research proposes code readability scores derived from linters or models to quantitatively evaluate the readability of generated code snippets~\cite{posnett2011simpler}.

{\bf Reliability} is the degree to which a system performs specified functions under specified conditions for a specified period.
Traditional methods involve fault injection testing, stress testing, and calculating metrics like mean time between failures~\cite{natella2016assessing,cotroneo2020fault}.
These methods require a long-running, stateful application and are thus unsuitable for evaluating stateless code snippets generated by LLMs.
Recent studies have explored using a powerful LLM as a judge to score the reliability of code snippets based on a predefined set of rules, such as evaluating whether the code snippet implements necessary handling for edge cases~\cite{zhang2025codecriticbench}.

Based on the existing methods, in this study, we implement a series of metrics to assess the quality of code snippets generated by LLMs using different third-party libraries, thereby evaluating whether a library can be effectively used in the AI-assisted development.
More details are in~\autoref{s:eval_metric}.


\section{Motivation}
\label{sec:moti}

\begin{figure*}
	\centering
	\includegraphics[width=\linewidth]{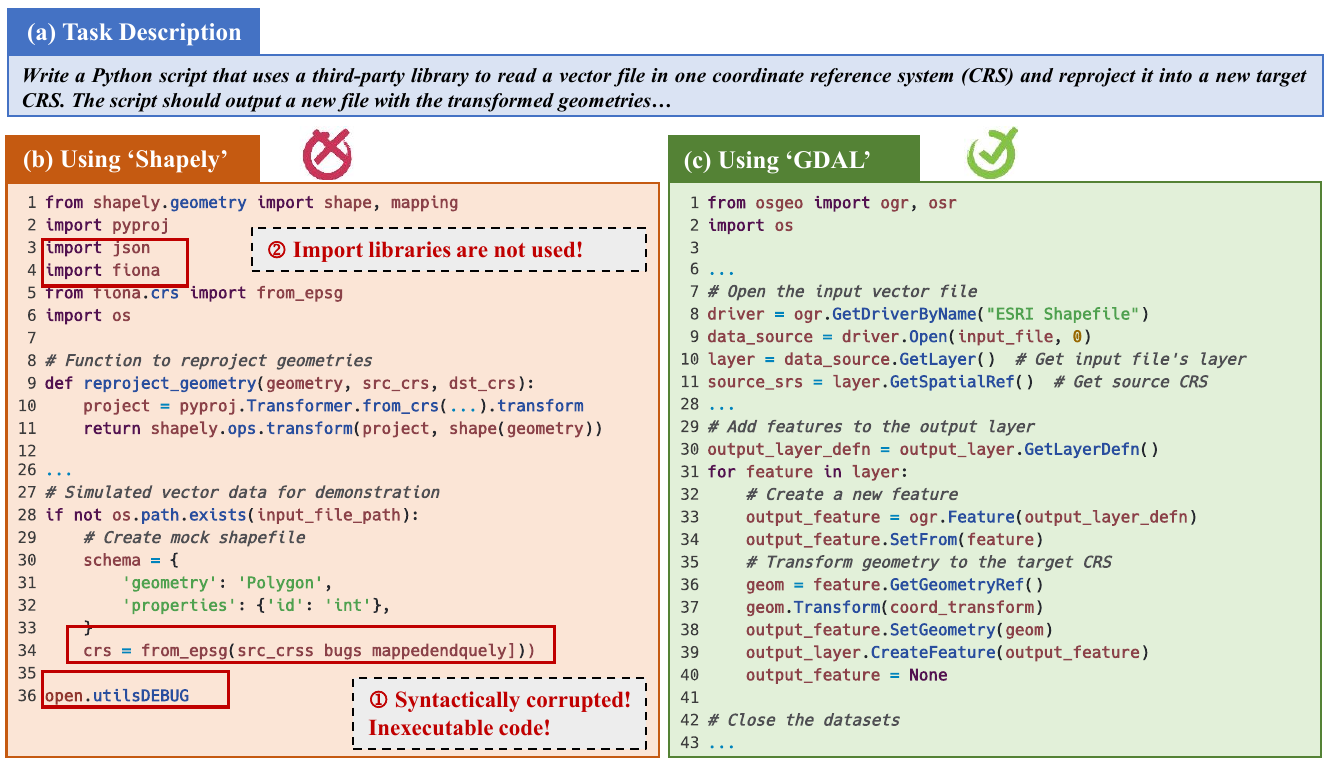}
	\caption{A Motivation Case on GPT-4o}
	\label{fig:moti}
\end{figure*}

\subsection{AI Coding Proficiency of Technology}
\label{s:moti_def}
As LLMs grow in influence across industry and daily life, people begin to pay attention to whether organizations and even the whole society are `ready' for the AI-powered production and lives~\cite{johnk2021ready,tehrani2024decoding,aireadiness}.
Inspired by this concept, in this paper, we introduce the \textit{AI coding proficiency of technology} as a new perspective for technology selection to answer the question `is a technology ready for AI-assisted development'.
Specifically, it evaluates whether the given model can effectively utilize the given technology to produce high-quality code snippets.
This measurement provides a parallel, complementary dimension to traditional technology selection frameworks and quality models that focus on intrinsic attributes of one technology.

\begin{definition}[AI coding proficiency of technology]
The AI coding proficiency of a technology refers to the degree to which a technology (e.g., a third-party library or service) can be effectively and skillfully utilized by a given AI model (e.g., LLMs) in software development processes.
Specifically, for technologies with high AI coding proficiency, an LLM can proficiently leverage their APIs and functions with minimal human intervention to produce high-quality code solutions that satisfy both functional requirements and critical non-functional attributes (e.g., maintainability and performance efficiency).
Higher AI coding proficiency translates to reduced integration and maintenance costs, lower technical debt, and more reliable outcomes in AI-assisted development workflows.
\end{definition}

Note that in this study, we assess the AI proficiency of a technology by measuring the quality of code snippets generated by LLMs when they are asked to use the given technology in code generation.
A high code quality score indicates high AI coding proficiency.

\subsection{Motivating Case}
\label{s:moti_case}



Two technologies with similar functionality and popularity can have significant differences in AI proficiency, introducing extra debugging costs and hidden risks when not considered AI proficiency in technology selection.
To motivate our study, \autoref{fig:moti} provides an example using GPT-4o.
We use two prompts to query the model with a task description under the `geographic data processing and analysis' scenario, as shown in~\autoref{fig:moti} (a), asking the LLM to generate Python scripts that read a vector file in a coordinate reference system (CRS) and reproject it to a new CRS.
The only difference is the requested library, \texttt{Shapely} (4.4k GitHub stars\footnote{\url{https://github.com/shapely/shapely}}) and \texttt{GDAL} (5.8k GitHub stars\footnote{\url{https://github.com/OSGeo/gdal}}), both for geographic data processing.
Despite similar functionality and popularity, we observe significant quality differences in the generated code.
As shown in~\autoref{fig:moti} (b), using \texttt{Shapely}, the LLM produces a code snippet with functional omissions and syntax errors (\ding{172} in~\autoref{fig:moti}) that make it inexecutable, while also lacking detailed comments and importing two unused libraries (\ding{173}), resulting in poor maintainability and increased debugging and maintenance costs.
Under our evaluation metrics (\autoref{s:eval_metric}), this snippet scores 30.00 for functional suitability and 49.13 for maintainability.
The overall score is only 48.83.
In contrast, as shown in~\autoref{fig:moti} (c), using \texttt{GDAL}, the LLM generates high-quality, executable code that provides clear comments and avoids unnecessary dependencies, achieving 100.00 for functional suitability and 86.40 for maintainability.
The overall score achieves 90.28, which exhibits an 84.89\% improvement over \texttt{Shapely}.
We provide the full prompts and the generated code snippets for this example in our repository to facilitate reproduction and review~\cite{our_repo}.


Notably, the significant difference in quality scores is not an isolated case or an outlier.
In repeated runs on this set of prompts, we consistently observe the same pattern of code quality differences.
Furthermore, our large-scale experiments on six LLMs show that 11.17\% of library pairs exhibit significant quality score gaps, underscoring the need to consider AI proficiency in technology selection in the LLM era.
Without such guidance, selecting a library with low AI proficiency directly erodes the productivity gains of AI-assisted development.
It pulls developers away from an efficient generative workflow into a tedious fix-and-correct loop, driving up debugging, refactoring, and integration costs.
More seriously, it incurs technical debt and may even introduce potential security risks to development.
From a broader perspective, AI proficiency also has profound implications for the software community and ecosystem.
Developers naturally prefer to adopt higher-quality code snippets.
As a result, low-proficiency libraries will gradually be marginalized in the competitive landscape of the LLM era.
This AI-driven `Matthew effect' could accelerate a winner-takes-all market dynamic, threatening healthy competition in the digital market and reducing the robustness and diversity of the entire software ecosystem.
Furthermore, excessive reliance on a small number of libraries with high proficiency in the software ecosystem can lead to a centralization of users, which increases the risk of known exposures being reused in attacks and amplifies the impact of targeted poisoning attacks~\cite{pearce2022asleep,decan2018impact}.

To systematically study the AI proficiency and explore its implications, we conduct an empirical study on six widely used LLMs (i.e., GPT-4o-mini, GPT-4o, Claude Sonnet 4, Gemini 2.5 Flash, Qwen3-Coder-Plus, and DeepSeek-R1) with a dataset covering 61 task scenarios and 170 third-party libraries.
From the experiments, we distill six key findings, aiming to provide new insights for software development and technology selection in the LLM era.

\section{Evaluation Methodology}
\label{sec:system}



\subsection{Dataset Construction}
\label{s:dataset}
To explore and study the AI proficiency of different third-party libraries on various task scenarios, we design a pipeline to construct a comprehensive dataset through the following two stages, namely \textit{scenario and library collection} and \textit{prompt generation}.

\noindent
\(\bullet\)
{\it Scenario and Library Collection.}
This stage begins with a collection of diverse software development task scenarios inspired by previous studies~\cite{zhang-etal-2025-invisible}. 
We then collect competing third-party libraries related to each scenario.
Note that the `competing libraries' here are defined as multiple libraries within the same scenario whose core functionalities highly overlap, enabling them to achieve equivalent goals under the same task description.
Our study focuses on Python code generation of LLMs, and all collected third-party libraries should be able to be used in Python (e.g., via Python adapters or bindings).

The analysis of collected libraries shows that most of the libraries are typically only applicable to specific tasks within a scenario, making it difficult for them to be considered as `competing libraries'.
Take the `Machine Learning' scenario as an example.
The collected libraries contain \texttt{scikit-learn}, \texttt{TensorFlow}, and \texttt{PyTorch}.
The \texttt{Scikit-learn} library is primarily used for traditional machine learning tasks (e.g., logistic regression, random forests), whereas \texttt{TensorFlow} and \texttt{PyTorch} are more commonly used for deep learning tasks (e.g., neural network training and inference).
Therefore, \texttt{scikit-learn} is not a direct competitor to \texttt{TensorFlow} and \texttt{PyTorch}.
In contrast, the functional overlap between \texttt{TensorFlow} and \texttt{PyTorch} is high, making them fit for the `competing libraries'.

To resolve this issue, we invite two experts with professional backgrounds in SE to further refine and restructure the scenarios based on the functional boundaries of the libraries, and to filter for `competing libraries' within each refined scenario.
For any disagreements between the two experts on the reconstruction results, a third expert is invited to facilitate a discussion to reach a consensus.
Any refined scenario with fewer than two competing libraries will be discarded.
Finally, we collect 170 libraries across 61 refined scenarios.
More details about scenarios and libraries are in~\cite{our_repo}.

\noindent
\(\bullet\)
{\it Prompt Generation.}
We build an automated prompt generation pipeline to generate prompts for third-party libraries.
Based on prior work~\cite{zhang-etal-2025-invisible}, our prompt template is designed as follows:

\begingroup
\addtolength\leftmargini{-25pt}
\begin{quote}
    \it Write a Python script that uses `\textbf{(\underline{LIBRARY})}' to complete the following coding task: `\textbf{(\underline{DESCRIPTION})}'.
\end{quote}
\endgroup

The \textbf{(\underline{LIBRARY})} indicates the name of the collected third-party library, and \textbf{(\underline{DESCRIPTION})} is a specific task description aligned with the scenario.
For example, in the `Data Visualization' scenario, \textbf{(\underline{DESCRIPTION})} could be `\textit{Create a line chart that plots the monthly sales data, displaying both the sales amount and the months on the appropriate axes}'.
To efficiently obtain task descriptions, the pipeline queries an LLM (i.e., GPT-4o) to generate five candidate descriptions for each scenario.
It then queries the model again to validate these candidates, ensuring that the given task can be accomplished by any competing library within this scenario.
For candidates that fail validation, the pipeline triggers regeneration until the retry budget is exhausted (up to three retries).
In total, this stage produces 850 prompts based on the collected third-party libraries.

\subsection{Evaluation Metrics}
\label{s:eval_metric}
Based on the evaluation methods in~\autoref{s:quality_eval}, this study uses the following metrics to score the code snippets generated by LLMs on five commonly used quality dimensions.
All scores are normalized to the range of \([0,100]\), and a higher value indicates better quality on that dimension.

\noindent \(\bullet\)
{\it Functional Suitability Score.} Following the prior work~\cite{tong2024codejudge,crupi2025effectiveness,zhang2025codecriticbench}, we use an LLM to review the code snippets and score their functional correctness in conjunction with the task descriptions.

\noindent \(\bullet\)
{\it Performance Score.} Following the prior work~\cite{wang2025can,zhang2025codecriticbench}, we use an LLM to estimate the time complexity and space complexity of the code snippets and take their mean as the performance score, which reflects the performance of LLM-generated code snippets.


\noindent \(\bullet\)
{\it Maintainability Score.} We calculate the widely used Maintainability Index (MI) as the maintainability score~\cite{oman1992metrics}.
This index is calculated using the Halstead volume, cyclomatic complexity, and the number of code lines.
A higher score indicates lower maintenance difficulty and stronger maintainability.

\noindent \(\bullet\)
{\it Usability Score.} We adopt a readability evaluation model from prior work to score the readability of each snippet as the Usability Score~\cite{posnett2011simpler}.
This model uses features such as Halstead volume and the number of non-empty lines to reflect how easy the code is to read and use.

\noindent \(\bullet\)
{\it Reliability Score.} Following prior work~\cite{wang2025can,zhang2025codecriticbench}, we use the LLM to assess the reliability of the given code snippets (e.g., whether exceptions and edge cases are handled appropriately).

Following the prior work~\cite{white2025livebench}, we compute the arithmetic mean of the five dimension scores as the overall code quality score \(S\), providing a more comprehensive view of the overall quality of the given code snippets.
Note that we adopt equal weights as a neutral evaluation to avoid introducing subjective bias into this study and discuss the future work of alternative weighting in~\autoref{sec:discuss}.
We define the AI proficiency score \(\mathcal{P}\) of the library \(l\) as the average overall quality scores of code snippets generated by the model \(m\) across all prompts involving \(l\) in the scenario \(s\), as shown follows:
$$
\mathcal{P}
= \frac{1}{\lvert P_{l,s}\rvert}
  \sum_{p \in P_{l,s}} S^m_{p},
$$
where \(P_{l,s}\) indicate the set of prompts in a given scenario \(s\) that ask the LLM to generate code snippets with library \(l\), and \(S^m_{p}\) indicates the overal quality score of the code snippet produced by model \(m\) for prompt \(p\).
\section{Experimental Results \& Analysis}
\label{sec:exp}

In this section, we report and analyze the experimental results to answer the following research questions (RQs):
\begin{enumerate}
    \item {\bf RQ1}: How is the AI proficiency of different libraries on different LLMs? 
    \item {\bf RQ2}: What typical failure patterns do LLMs exhibit when generating low-quality code snippets? 
    \item {\bf RQ3}: Can existing prompting techniques improve the AI proficiency of libraries? 
\end{enumerate}

\subsection{Setup}
\label{s:model}
In this study, we select six LLMs that cover mainstream closed-source and open-source models and have a large user base.
Specifically, 
\ding{182} {\bf GPT-4o \& GPT-4o-mini} are closed-source models developed by OpenAI.
Both of them have powerful reasoning capabilities and act as the backbones of ChatGPT~\cite{gpt4o,gpt4omini}.
We use `GPT-4o-2024-11-20' and `GPT-4o-mini-2024-07-18' in the experiments.
\ding{183} {\bf Claude Sonnet 4} is a closed-source commercial model developed by Anthropic.
It exhibits powerful code generation capabilities and serves millions of users monthly~\cite{claude4,claude4users}.
We use `Claude Sonnet 4-20250514' in the experiments.
\ding{184} {\bf Gemini 2.5 Flash}, developed by Google, has a context window of over one million tokens and powerful reasoning and coding capabilities~\cite{gemini25}.
\ding{185} {\bf Qwen3-Coder-Plus} is the latest generation of large code-specific models developed by Alibaba~\cite{qwen3}.
We use `Qwen3-Coder-Plus-2025-07-22' in the experiments.
\ding{186} {\bf DeepSeek-R1} is an open-source LLM developed by DeepSeek~\cite{guo2025deepseek}.
DeepSeek-R1 is currently one of the most powerful open-source LLMs with state-of-the-art coding capabilities.

According to the experimental results of the existing benchmarks~\cite{leaderboard1,yang2025qwen3}, the coding capabilities of these models are ranked as follows (from high to low), DeepSeek-R1, Claude Sonnet 4, Qwen3-Coder-Plus, Gemini 2.5 Flash, GPT-4o, and GPT-4o-mini.
During the experiments, we follow the official documentation of each model and use default parameter configurations.
We repeatedly query each prompt five times on each model and ultimately collect 25,500 LLM responses from the six models that completed the given task description using the target third-party libraries.

In addition, to enhance the robustness and reliability of our analysis in experiments, we employ a widely used bootstrap resampling strategy.
When calculating the AI proficiency score, we use the bootstrap method to resample 1,000 samples of LLM responses, and then calculate the mean and 95\% confidence interval~\cite{mooney1993bootstrapping,deldjoo2024understanding}.
The significance of the experimental results and analysis has been statistically tested (\eg, t-test).
Based on the results of prior work~\cite{zhang2025codecriticbench}, the LLM-based metrics in this study use the `O1-mini-2024-09-12' model developed by OpenAI to efficiently obtain high-quality evaluation results.

\subsection{RQ1: Proficiency Gap}
\label{s:rq1}



\begin{figure}
	\centering
	\includegraphics[width=\linewidth]{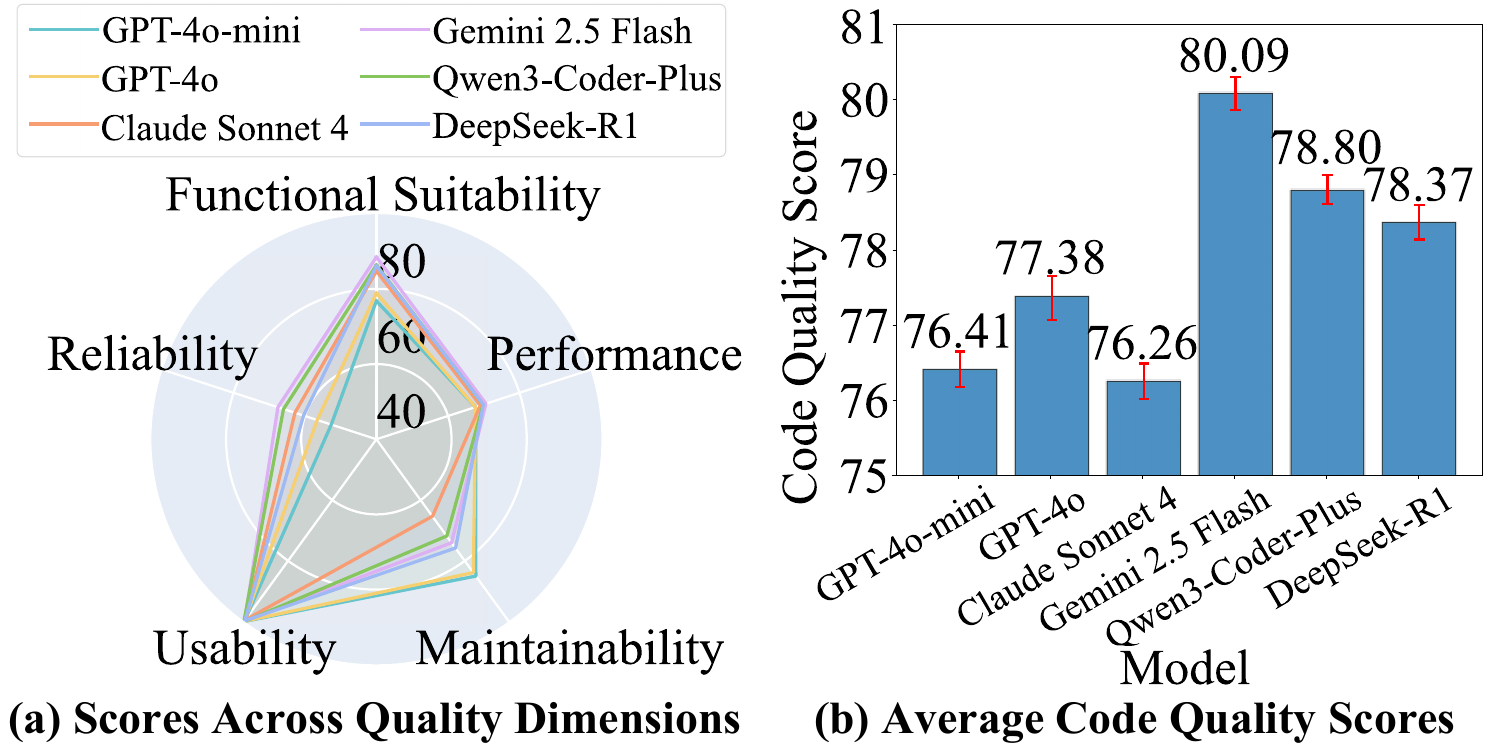}
	\caption{Comparison of Code Quality Across LLMs}
	\label{fig:rq1_model}
\end{figure}



To systematically study the AI proficiency of different libraries across models, we analyze a total of 25,500 responses from six LLMs.
For each model response, we compute the quality score of its code snippets using the metrics in~\autoref{s:eval_metric}.
We then conduct two complementary comparisons.
\ding{182} \textit{Model-level.}
We first compute each model's average quality score over all its responses to characterize its overall proficiency and rank the models accordingly.
We also analyze the correlation between this proficiency ranking and the ranking of models' general coding capability.
Next, for each library, we compare its AI proficiency scores across different models to identify which model the library achieves the best proficiency on and obtains the highest code quality scores, aiming to provide empirical guidance for model selection for the AI-assisted development.
\ding{183} \textit{Library-level.}
Based on the library information collected in~\autoref{s:dataset}, we first pair functionally similar libraries and then quantify the differences between their code quality across scenarios with Cohen's d effect size (threshold is 0.5~\cite{cohen2013statistical}).
This allows a fine-grained understanding of the proficiency gaps between competing libraries, aiming to provide empirical guidance for technology selection in the LLM era.



\noindent
\(\bullet\)
{\bf Comparison between LLMs.}
The scores of the code snippets generated by each LLM on the five quality dimensions are shown in~\autoref{fig:rq1_model} (a).
We can observe that LLMs generally obtain high scores on the usability dimension, as most of the generated code snippets are well-commented and clearly structured, which is conducive to understanding and reuse.
On other dimensions, LLMs perform differently.
For example, GPT-4o often produces smaller, lower-complexity, and more maintainable code snippets in our dataset (an average MI of 83.82), but these code snippets are more prone to functional errors and provide weaker coverage of edge cases and exception handling, leading to lower scores on functional suitability and reliability.
In contrast, Claude Sonnet 4 tends to deliver functionally correct responses that align well with the task description, but these code snippets include extensive parameter checks and try-except logic, resulting in a low MI value (i.e., 65.32).

\autoref{fig:rq1_model} (b) reports the overall code quality scores of different models, and red marks the 95\% bootstrap confidence intervals.
Gemini 2.5 Flash achieves the highest score, indicating stronger AI proficiency across the 170 libraries of our dataset.
It is followed by Qwen3-Coder-Plus and DeepSeek-R1.
Affected by the low MI values, Claude Sonnet 4 has the lowest code quality score among these LLMs.
Furthermore, we use Spearman's rank correlation coefficient~\cite{sedgwick2014spearman,gauthier2001detecting} to analyze the correlation between the code quality ranking and the ranking of LLMs' coding capability (\autoref{s:model}).
The Spearman coefficient is 0.09, indicating no significant correlation between the two rankings.
Note that the rankings in functional suitability, performance, and reliability show a moderate correlation with model capability, but are still not statistically significant.
This discrepancy may stem from the fact that existing benchmarks emphasize correctness while paying little attention to other important code quality dimensions like maintainability and reliability.
As a result, the models cannot effectively learn quality-related knowledge during training, which highlights the limitations of current evaluation practices.
Given that millions of developers now integrate LLMs into their development workflows, we argue that selecting an assisted-development model should not merely rely on correctness scores.
Multi-dimensional quality metrics should also be considered to identify models that can truly reduce cost and improve development efficiency.

\begin{figure}
	\centering
	\includegraphics[width=\linewidth]{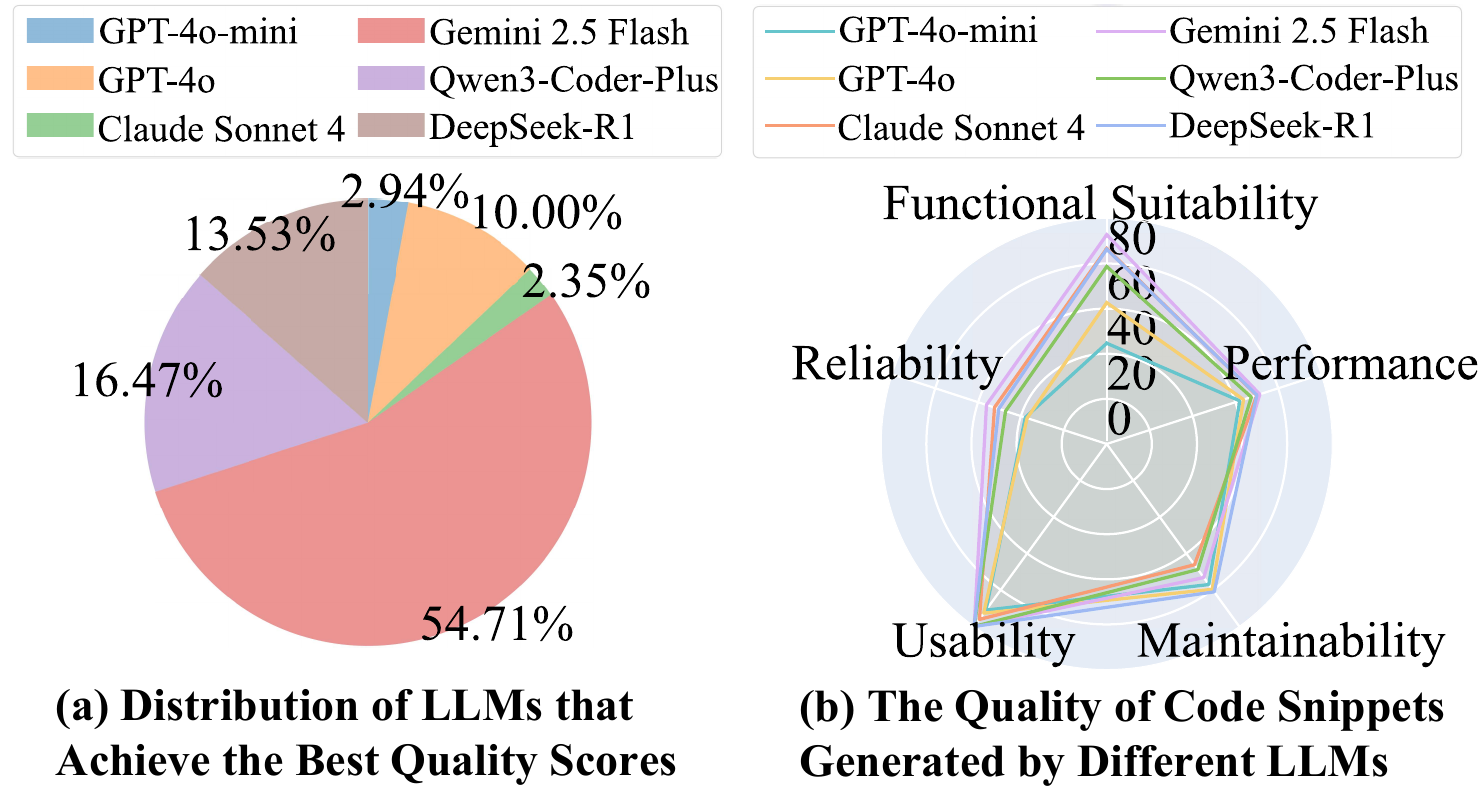}
	\caption{Analysis of AI Proficiency Scores Across LLMs}
	\label{fig:rq1_prof}
\end{figure}

\finding{Gemini 2.5 Flash achieves the highest code quality score among six models, while Claude Sonnet 4 achieves the lowest quality score. In addition, the ranking of LLMs by overall code quality scores shows no significant correlation with the coding capability ranking.}

%

For each library, we evaluate the differences between its AI proficiency scores across different LLMs.
As shown in~\autoref{fig:rq1_prof} (a), the models achieving the highest proficiency scores are unevenly distributed across the libraries.
Gemini 2.5 Flash achieves the highest score on 93 libraries, ahead of other models, whereas Claude Sonnet 4 ranks first on only 4 libraries.
Our analysis shows that there is no significant correlation between the model ranking using these library counts and the model capability ranking (the Spearman coefficient is 0.09).
Our analysis further shows that the quality score of code snippets using the same library can vary substantially across LLMs (up to 16.17 in~\autoref{fig:rq1_prof} (b)).
In this case, the prompt asks models to perform 3D modeling tasks using \texttt{Three.js} and its Python binding \texttt{pythreejs}.
However, in repeated runs, GPT-4o-mini often fails to invoke the correct APIs of these libraries, leading to inexecutable and incorrect code snippets with a low proficiency score.
Such frequent errors may stem from limited exposure or insufficient learning of the relevant APIs and documentation during training, resulting in poor code quality in practical coding tasks.

From an engineering perspective, low AI proficiency could directly erode the productivity gains of AI-assisted development, increase debugging and integration costs, and introduce technical debt and quality risks.
Therefore, for technologies that must be used in practice, developers should prioritize models that demonstrate proficiency in the target technology to accelerate development and ensure delivery quality.

\finding{The best choice of LLMs with the highest proficiency score varies across libraries.
Gemini 2.5 Flash can achieve the highest score on the most libraries (93/170), while GPT-4o-mini and Claude Sonnet 4 lead on only 4 libraries.
It underscores that the model choice in AI-assisted development should be technology-specific, guided by the AI proficiency of the target technology, rather than relying solely on the leaderboards of general coding capability.}


\begin{table}
\caption{Competitive Libraries With Significant Quality Differences Across Six LLMs}
\label{tab:rq1}
\centering
\footnotesize
\tabcolsep=5pt
\begin{tabular}{ccc}
\toprule
Model Name & \#Pair & Ratio (\%) \\
\midrule
GPT-4o-mini       & 23 & 12.23 \\
GPT-4o            & 28 & 14.89 \\
Claude Sonnet 4   & 20 & 10.64 \\
Gemini 2.5 Flash  & 17 &  9,04 \\
Qwen3-Coder-Plus  & 16 &  8.51 \\
DeepSeek-R1       & 22 & 11.70 \\
\midrule
Total             & 126 & 11.17 \\
\bottomrule
\end{tabular}
\end{table}

\begin{figure}
	\centering
	\includegraphics[width=\linewidth]{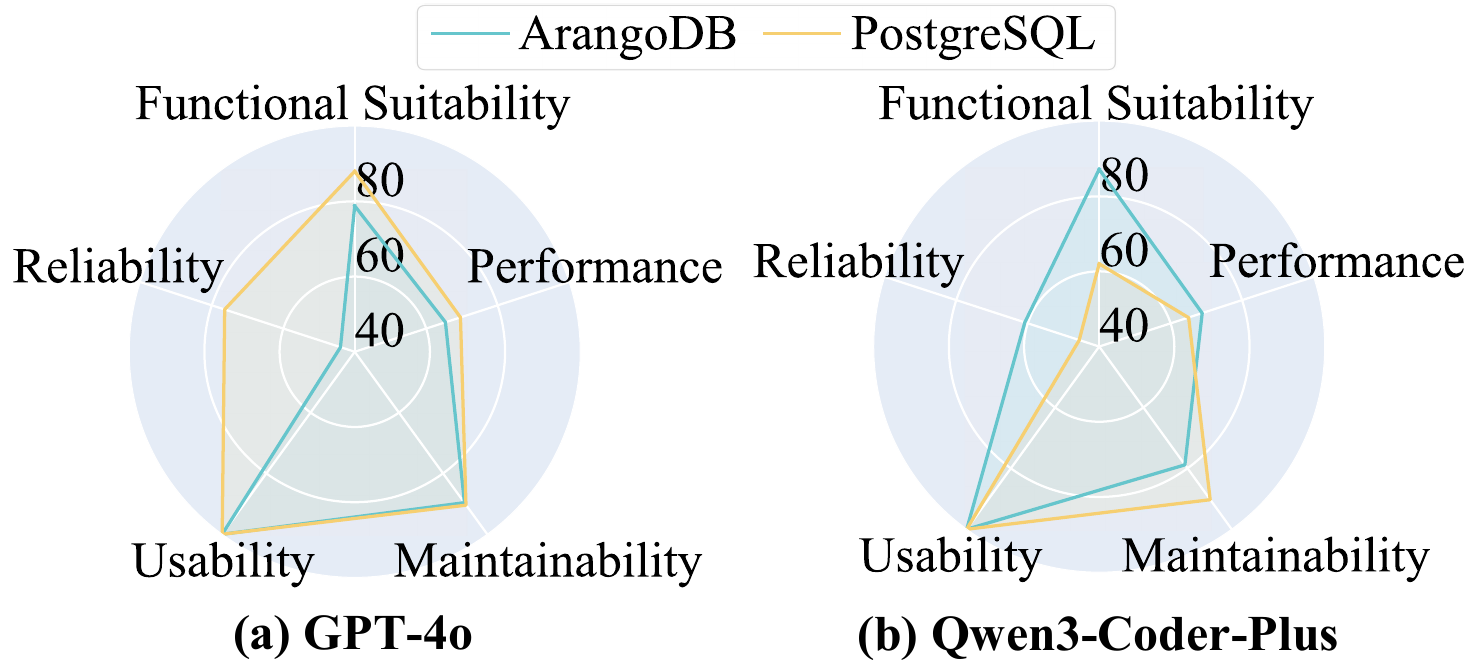}
	\caption{Code Quality Scores of Competing Libraries on Two LLMs}
	\label{fig:rq1-competing-case}
\end{figure}

\smallskip
\noindent
\(\bullet\)
{\bf Comparison between Libraries.}
For third-party libraries that have similar functionalities, we pair them to form 188 competing library pairs.
We then compare the quality scores of each pair under the same scenario and use Cohen's effect size to identify pairs with significant quality differences.
\autoref{tab:rq1} shows the number and ratio of competing pairs flagged as `Significant Difference' across six LLMs, highlighting the prevalence of substantial gaps in AI proficiency among functionally similar libraries.
In addition, seven pairs of libraries (e.g., the pair of Python adapters for \texttt{CouchDB} and \texttt{PostgreSQL}) show significant differences on at least three LLMs.
Our further analysis shows that 86.17\% of competing pairs exhibit different winners across models (i.e., the leading library changes with the model).
\autoref{fig:rq1-competing-case} provides an example.
When GPT-4o and Qwen3-Coder-Plus are asked to complete tasks related to the database using the Python adapters of \texttt{ArangoDB} and \texttt{PostgreSQL}, GPT-4o achieves larger code quality scores with \texttt{PostgreSQL} than with \texttt{ArangoDB}, while Qwen3-Coder-Plus achieves better code quality using the \texttt{ArangoDB} library.
Such different leadership on different LLMs indicates that, for technology selection, comparing only the libraries or only the models is insufficient.
There is an urgent need to introduce and quantify the AI proficiency dimension of technologies to ensure that the selected technologies can produce high-quality and stable outputs in specific scenarios and given models.

\finding{11.17\% of competing library pairs show significant differences in the experiment.
Among competing library pairs, which one delivers better code quality and proficiency depends largely on the selected model, further highlighting the importance of incorporating AI proficiency into the technology selection in the LLM era.}

\subsection{RQ2: Failure Pattern}
\label{s:rq2}

To gain a deeper understanding of the manifestations of low AI proficiency and its profound impact on software development, we analyze the code snippets with low quality scores and construct a failure pattern taxonomy based on this analysis.
Specifically, we first rank all code snippets from small to large based on their code quality scores and then use the interquartile range method (threshold is \(Q_1-1.5*IQR\))~\cite{tukey1977exploratory} to identify the samples with the low quality scores among all 25,500 code snippets, resulting in a total of 814 low-scoring code snippets.
We then invite two SE experts to annotate the failure patterns in these code snippets.
Each expert has at least five years of Python development experience.
Before annotation, both experts receive unified training to familiarize themselves with the scoring methods for each code quality dimension in the~\autoref{s:eval_metric}.
During the annotation process, annotators can access the source code, the quality scores for each dimension, and the supporting reasons for the relevant quality scores from the LLM evaluator.
The Cohen's Kappa coefficient of these annotation results reaches 0.85.
Then, a third expert facilitates a discussion to solve any disagreements, ultimately reaching a consensus on the annotation results and developing a stable failure pattern taxonomy.
Note that we apply a multi-label annotation as a single code snippet can exhibit multiple patterns simultaneously (e.g.,~\autoref{fig:moti} (b)).

\noindent
\(\bullet\)
{\bf Taxonomy of Failure Patterns.}
Based on the annotation results, \autoref{tab:rq2-taxonomy} summarizes eight categories of failure patterns (P1-P8), and we introduce them as follows:

\textit{P1: Syntax Error.}
This category of pattern occurs when a generated code snippet is not executable due to syntax errors or incomplete snippets (e.g., mismatched indentation and truncated statements).
In \autoref{s:moti_case} (b), an extra `])' appears on Line 34, causing the code snippet to fail syntax checking.
Such errors force developers to engage in iterations of locating, fixing, and rewriting, which not only prolongs debugging and integration processes but can also increase overall delivery costs and schedule risks.
In addition, frequent naive errors may force developers to abandon the use of specific models or libraries, potentially affecting the digital market and software ecology.

\textit{P2: Incorrect Library Usage.}
This failure pattern occurs when a snippet contains the incorrect usage of a target library or its APIs, for example, invoking deprecated or non-existent functions.
The code snippets may also select APIs ill-suited to the input characteristics or fail to call the target library as required.
Compared with syntax errors, such a misuse is more insidious.
It could lead to logical errors that are difficult to identify, posing security and reliability risks to software development.

\textit{P3: Incorrect Functionality.}
In this pattern, the code snippet generated by the LLM may be executable, but it implements incorrect algorithms and fails to meet the core task requirements.
For instance, when asked to build a convolutional neural network in PyTorch, the model (e.g., Qwen3-Coder-Plus) may return a classifier without convolutional layers, which is undoubtedly not enough for the task.
This pattern is the most common across six LLMs, reflecting the limitations of LLMs in performing complex tasks relevant to production scenarios to a certain extent.
Frequent functionality issues could force the developers to repeatedly rework and verify, erode trust in automated coding assistance, and increase testing and review overhead across the pipeline.

\textit{P4: Missing Edge Handling.}
The code snippet omits checks or handlers for edge cases and inputs (e.g., empty inputs and extreme values), making it less reliable and likely to fail under certain conditions.
Such failures are often expensive to troubleshoot and difficult to expose in tests, posing a threat to the security and reliability of software development.

\textit{P5: Inefficient Implementation.}
This pattern is related to the performance efficiency dimension.
It occurs when the code snippet contains inefficient implementations such as redundant and frequent I/O, nested loops, and exception checking.
This directly leads to higher latency and resource consumption, which degrades user experience and software scalability and accrues as performance debt, making later optimization riskier and more expensive.

\textit{P6: Missing Comments.}
The code snippet provides insufficient documentation for complex classes or key steps.
Missing or minimal comments will increase cognitive load and misuse risk and slow code review, harming the long-term usability of the code snippet. 

\textit{P7: Import Issues.}
This failure pattern includes the situations where code snippets use APIs from specific libraries but do not import them, or import modules that are never used.
These issues can cause execution failures, pollute the environment, and cause dependency bloat, complicating software deployment and leading to potential issues in subsequent version control and dependency.

\textit{P8: Others.}
This category includes various code quality problems, e.g., excessive demo inputs, which depress the scores in dimensions like maintainability.
Such issues can hinder human developers from understanding the code snippets generated by LLMs, slow down their learning speed, thereby reducing project maintainability and increasing development costs.

\begin{table}
\caption{Taxonomy of Failure Patterns}
\label{tab:rq2-taxonomy}
\centering
\tabcolsep=1.5pt
\footnotesize
\begin{tabularx}{\linewidth}{c c X}
\toprule
No. & Failure Pattern & \multicolumn{1}{c}{Description} \\
\midrule
P1 & Syntax Error & Containing syntax errors and missing content, making the code snippet inexecutable. \\
P2 & Incorrect Library Usage & Containing incorrect usage of the libraries and corresponding APIs. \\
P3 & Incorrect Functionality & Containing incorrect implementation, resulting in a failure to meet the core requirements of the given task. \\
P4 & Missing Edge Handling & Failing to consider or process the edge cases, which may still cause errors on specific inputs. \\
P5 & Inefficient Implementation & Containing inefficient implementations, which can still be optimized. \\
P6 & Missing Comments & Lacking necessary comments and explanations for complex functions, classes, or key steps. \\
P7 & Import Issues & Using libraries that are not imported, or the imported library is not used. \\
P8 & Others & Other issues, such as containing too many demo inputs, leading to a low maintainability index. \\
\bottomrule
\end{tabularx}
\end{table}


\begin{table}
\caption{Distribution of Failure Cases}
\label{tab:rq2-result}
\centering
\tabcolsep=2pt
\footnotesize
\begin{tabular}{cccccccccc}
\toprule
Model & P1 & P2 & P3 & P4 & P5 & P6 & P7 & P8 & Total \\
\midrule
GPT-4o-mini       & 15  & 31  &  89  & 81 & 73  & 50  & 34  &  2 & 375 \\
GPT-4o            & 177 & 42  & 154 & 58  & 37  & 15  & 25  &  0 & 508 \\
Claude Sonnet 4   & 51  & 19  & 94  & 77  & 62  & 62  & 31  & 12 & 408 \\
Gemini 2.5 Flash  & 10  & 20  & 32  & 41  & 29  & 24  &  7  & 16 & 179 \\
Qwen3-Coder-Plus  &  1  &  5  & 28  & 29  & 29  &  4  & 10  &  2 & 108 \\
DeepSeek-R1       & 25  & 11  & 35  & 45  & 36  & 28  & 13  &  5 & 198 \\
\midrule
Total & 279 & 128 & 432 & 331 & 266 & 183 & 120 & 37 & - \\
\bottomrule
\end{tabular}
\end{table}

\noindent
\(\bullet\)
{\bf Analysis of Failure Patterns.}
\autoref{tab:rq2-result} shows the distribution of failure cases across six LLMs.
Our analysis shows that `Incorrect Functionality' and `Missing Edge Handling' are the two most common failure patterns, separately accounting for 53.07\% and 40.66\% of all failure cases.
This finding is alarming, as these two types of issues are closely related to the quality and output of software development.
Functional correctness is the cornerstone of software, while comprehensive edge case and exception handling are key to ensuring stable and reliable system operation in real-world industrial environments~\cite{de2017revisiting,de2018studying}.
When models systematically ignore these edge handling processes, they effectively plant a `ticking time bomb' in the code snippets.
It could lead to various hidden dangers during application deployment, ranging from security vulnerabilities caused by insufficient input validation to service crashes caused by uncaught exceptions.
Therefore, one of the most serious dangers exposed by the low AI proficiency of a library is that it induces models to generate code snippets with inherent defects in functional correctness and reliability, thereby amplifying these security risks in the LLM era.

In addition, the pattern of `Syntax Errors' accounts for a significant proportion of GPT-4o's outputs.
Among the 4,250 code snippets generated by GPT-4o, 250 of them contain syntax errors, accounting for 5.88\%.
These seemingly naive errors can directly lead to build or integration failures in real-world engineering processes, significantly extending debugging cycles and severely undermining development teams' trust in AI-assisted development.
For model providers, persistent syntax issues not only mask deeper product flaws (e.g., low-efficiency code) but also undermine the model's iteration efficiency, user experience, and market reputation.

\finding{The most common failure patterns of low-quality code snippets are `Incorrect Functionality' and `Missing Edge Handling'.
These can seriously endanger the correctness and reliability of software systems and introduce potential security issues in software development.
It highlights the gap between LLM code generation results and the requirements of actual software development scenarios.}

\subsection{RQ3: Prompting Techniques}
\label{s:rq3}

To explore the impact of prompting techniques on AI proficiency of libraries, especially their effect on code quality gaps between competing libraries, we implement and evaluate three general prompting methods in this section.
\ding{182} {\bf CoT} is the zero-shot variant of Chain-of-Thought prompting, which includes the phrase `Let's think step by step' in the system prompt~\cite{kojima2022large}.
It can encourage the model to think gradually and generate structured and detailed answers.
Prior work has demonstrated that CoT can improve performance on reasoning and coding tasks~\cite{wei2022chain,kojima2022large,li2025structured}.
\ding{183} {\bf Few-shot prompting} provides examples in the prompt and guides the model to generate high-quality answers.
Following the prior work~\cite{xu2024does,li2023large,promptHub2025fewShotGuide,li2025structured}, for each prompt with a target library and a specific task, we provide a dedicated example containing a different task and a corresponding code snippet using the target library (overall score exceeding 80).
\ding{184} {\bf Regeneration} simulates the user workflow of `regenerating when dissatisfied'.
For each prompt, the model independently produces three candidate responses.
We then select the one with the highest overall quality score as the final output.

We evaluate the above three methods on six LLMs.
Due to resource constraints, we randomly select 20 pairs of prompts for each model.
Each pair contains two competing libraries under the same task.
10 pairs exhibit a significant quality gap between their original generated code snippets without the prompting methods, whereas the other 10 pairs do not (Cohen's effect size).
Following the setup in~\autoref{s:model}, we repeatedly query each prompt five times on each model.
\autoref{tab:rq4} shows the experimental results across six LLMs, where `Score' represents the average overall code quality scores across LLMs, `Gap' indicate the absolute difference in overall code quality scores between competing library pairs, and `Cons.' represents the proportion of pairs whose leading relationship of code quality scores between two libraries remains unchanged after using different prompting methods.
`Origin' represents the results without using any prompting method.
`Regen.' is the abbreviation of the regeneration method.

\noindent
\(\bullet\)
{\bf Analysis of Code Quality Score.}
As shown in~\autoref{tab:rq4}, prompt techniques can significantly improve the quality of code snippets generated by LLMs.
Specifically, `Few-shot' and `Regeneration' significantly improve the overall quality score (Wilcoxon signed-rank test, \(p < 0.05\)), with the largest improvement reaching 5.38\% (`Few-shot' on GPT-4o).
In addition, both methods significantly improve the scores on other quality dimensions such as functional suitability and Reliability.
In contrast, `CoT' only provides slight gains on several models, and the improvement is small (the improved score is less than 1).
This may be because the advantage of `CoT' primarily lies in `explicit reasoning', while this study focuses more on quality dimensions such as the correctness and maintainability of implementation.
In such a situation, additional thought chains and reasoning may not improve the code quality but could lead to more complex implementations, increasing complexity and compromising maintainability scores, ultimately resulting in little improvement.
Developers in the LLM era should choose appropriate prompting techniques based on the actual task requirements to reach high AI proficiency and improve development productivity.

\finding{`Few-shot' and `Regeneration' significantly improve overall code quality and AI proficiency on multiple LLMs, while `CoT' is less effective in improving code quality.}

\begin{table*}
\caption{Impact of Prompting Techniques on Competing Libraries}
\label{tab:rq4}
\centering
\tabcolsep=4pt
\footnotesize
\begin{tabular}{ccccccccccccccccccc}
\toprule
\multirow{3}{*}{Method}
& \multicolumn{3}{c}{GPT-4o-mini} 
& \multicolumn{3}{c}{GPT-4o} 
& \multicolumn{3}{c}{Claude Sonnet 4} 
& \multicolumn{3}{c}{Gemini 2.5 Flash} 
& \multicolumn{3}{c}{Qwen3-Coder-Plus} 
& \multicolumn{3}{c}{DeepSeek-R1} \\
\cmidrule(lr){2-4} \cmidrule(lr){5-7} \cmidrule(lr){8-10} \cmidrule(lr){11-13} \cmidrule(lr){14-16} \cmidrule(lr){17-19}
& Score & Gap &  Cons. (\%)
& Score & Gap &  Cons. (\%)
& Score & Gap &  Cons. (\%)
& Score & Gap &  Cons. (\%)
& Score & Gap &  Cons. (\%)
& Score & Gap &  Cons. (\%)\\
\midrule
Origin  & 74.08 & 6.23 & -   & 77.01 & 8.47 & -   & 78.07 & 4.72 & -   & 81.06 & 4.10 & -   & 77.42 & 5.96 & -   & 78.87 & 5.68 & - \\
CoT & 73.44 & 3.77 & 65.00 & 77.87 & 5.48 & 55.00 & 77.59 & 2.78 & 60.00 & 80.85 & 2.43 & 55.00 & 78.00 & 4.05 & 75.00 & 79.07 & 3.77 & 45.00 \\
Few-shot& 75.90 & 5.04 & 50.00 & 81.15 & 4.97 & 45.00 & 80.04 & 2.82 & 30.00 & 82.41 & 3.33 & 50.00 & 80.35 & 3.28 & 70.00 & 81.15 & 4.27 & 60.00 \\
Regen.  & 77.08 & 6.00 & 75.00 & 80.82 & 6.30 & 85.00 & 79.76 & 4.69 & 95.00 & 82.68 & 4.50 & 85.00 & 79.86 & 5.45 & 85.00 & 80.59 & 5.31 & 100.00 \\
\bottomrule
\end{tabular}
\end{table*}


\noindent
\(\bullet\)
{\bf Analysis of Competing Libraries.}
We calculate the average code quality differences (absolute gaps) for competing library pairs, as shown in~\autoref{tab:rq4}.
We can observe that both `CoT' and `Few-shot' methods significantly reduce the quality gap between libraries (\(p < 0.05\)).
The `Few-shot' method, in particular, reduces the average gap by up to 44.93\% on Qwen3-Coder-Plus.
This provides a viable path to mitigate the potential `winner-takes-all' problem (Matthew effect) caused by differences in AI proficiency.
It also suggests that in the LLM era, when selecting a technology, prompting techniques can be used as an auxiliary means to regulate the AI proficiency of target libraries.
Note that the impact of prompting techniques is conditional.
For competing libraries with significant code quality gaps without prompting, all three prompting methods are effective in narrowing the gap.
However, for pairs with a narrow gap, applying prompting methods can actually lead to a `lead reversal' (where the score of library A is better initially, but library B achieves a higher score after prompting) and even widen the gap.
This phenomenon is most pronounced for the `Few-shot' method, which is highly dependent on the content and coverage of the selected examples.
When examples better align with a library's API conventions or usage patterns, they can greatly improve the quality of code generated by LLMs using that library, thereby changing the relative advantages of two libraries.

\finding{Prompt techniques such as `Few-shot' and `CoT' can narrow the AI proficiency gap between competing libraries.
However, for pairs with narrow gaps, prompting methods may lead to a shift in the leading relationship between the scores of libraries, or even widen the gap.}


\section{Discussion}\label{sec:discuss}

\noindent
\(\bullet\)
{\bf AI Proficiency and Popularity.}
A natural hypothesis is that more popular libraries should exhibit higher AI proficiency, as their abundant code examples and documentation increase the likelihood of comprehensive representation in LLM training corpora.
To test this hypothesis, we use GitHub stars as a proxy for library popularity and apply exact two-sided binomial tests on competing library pairs from~\autoref{s:rq1}.
The concordance rate between popularity and AI proficiency rankings is 0.574 (95\% CI \([0.503, 0.643]\), \(p < 0.05\)), with a very small effect size (Cohen's d = 0.15), indicating only a weak positive correlation.
Notably, among the 80 discordant competing pairs, the more popular library actually receives a lower AI proficiency score.
For example, \texttt{DVC} has 2.12 times more GitHub stars than \texttt{Tensorboard}, yet its AI proficiency score on GPT-4o is 6.46\% lower.
This divergence may be attributed to multiple factors, including heterogeneous quality of community use cases across libraries, data filtering strategies applied during LLM pre-training, and frequent API changes introduced by rapid release cycles.
These results indicate that popularity is an unreliable predictor of AI proficiency, reinforcing the necessity of treating AI proficiency as an independent dimension in technology selection.

\noindent
\(\bullet\)
{\bf Threat to Validity and Future Directions.}
\ding{182} \textit{Scope of technologies and languages.}
This study focuses on the Python ecosystem and third-party libraries, where LLMs have demonstrated state-of-the-art code generation performance~\cite{twist2025llms,qing2025effibench}.
Although the underlying methodology for dataset construction and quality evaluation is fundamentally language-agnostic, the current findings may not directly generalize to other programming languages or technology types (e.g., platforms and system-level tools).
Future work could develop cross-language (e.g., C++ and Java) and cross-platform benchmarks that encompass broader technology types and incorporate multi-technology collaboration tasks to better approximate real-world production environments.
\ding{183} \textit{Technology selection framework.}
This study examines AI coding proficiency as a single new dimension without integrating it with other emerging factors in AI-assisted development, such as AI efficiency (i.e., the token and time cost for generating satisfactory code) and agent proficiency (i.e., the readiness of a technology for invocation by agentic AI systems).
Moreover, the proficiency score in this study adopts equal weights across the five dimensions as a neutral setting, whereas real-world scenarios may require domain-specific weighting (e.g., prioritizing reliability over readability in safety-critical systems), which could affect the relative ranking of competing libraries.
Future research could extend established methodologies (e.g., ATAM~\cite{kazman2000atam}) to build a holistic technology selection framework that integrates AI coding proficiency with traditional quality attributes, designs adaptive dimension weighting strategies tailored to specific project requirements, and systematically balances these factors alongside other emerging dimensions.
\ding{184} \textit{Enhancement strategies.}
While~\autoref{s:rq3} demonstrates that prompting techniques can improve AI proficiency scores and narrow proficiency gaps, this exploration is limited to general-purpose methods applied from the user side.
Future work could explore other mitigation strategies from both the model side (e.g., targeted data augmentation and comparative fine-tuning to reduce proficiency disparities) and the library side (e.g., curating high-quality usage examples to facilitate effective learning).
\section{Conclusion}
\label{sec:conclusion}

This paper introduces \textit{AI coding proficiency}, a new perspective for technology selection in the LLM era that evaluates whether LLMs can effectively utilize a given technology to produce high-quality code snippets.
Through the large-scale empirical study on the AI proficiency of third-party libraries across six LLMs, we reveal that competing libraries with similar functionality can exhibit significant proficiency gaps on the same model, while the same library can also perform differently across models.
Low AI coding proficiency undermines the efficiency of AI-assisted development and can drive users toward high-proficiency libraries, exacerbating market concentration and harming the technological diversity of the software ecosystem.
We call on the community to incorporate AI proficiency into technology selection and take measures to ensure development quality and maintain an open, diverse software ecosystem.

%
\section{Data Availability Statement}
\label{sec:data}
To follow the Open Science Policy and support reproducibility, our pipeline, dataset, and necessary results are released at~\cite{our_repo}.

\newpage

\newpage

{ \scriptsize
\bibliographystyle{ACM-Reference-Format}
\bibliography{reference}
}

\end{document}